\documentstyle[12pt]{article}
\def\vcb{\mid V_{cb} \mid}
\def\vtd{\mid V_{td} \mid}
\def\vub{\mid V_{ub}/V_{cb} \mid}
\def\f{\frac}
\def\o{\over}
\def\b{\begin{equation}}
\def\e{\end{equation}}
\def\l{\label}
\def\kpnn{K^+\rightarrow\pi^+\nu\bar\nu }
\def\kpn{K^+\rightarrow\pi^+\nu\bar\nu}
\def\klpnn{K_L\rightarrow\pi^0\nu\bar\nu}
\def\klpn{K_L\rightarrow\pi^0\nu\bar\nu}

\def\imlt{{\rm Im}\lambda_t}
\def\relt{{\rm Re}\lambda_t}
\def\relc{{\rm Re}\lambda_c}

\begin{document}
\thispagestyle{empty}
\begin{flushright}
 MPI-PhT/94-30 \\
 TUM-T31-64/94 \\
 May 1994
\end{flushright}
\vskip1truecm
\centerline{\Large\bf  CP Violation: Present and Future
   \footnote[1]{\noindent Invited talk given at the First International
 Conference on
Phenomenology of Unification from Present to Future, Rome -
March 23-26, 1994, to appear in the proceedings.\\
   Supported by the German
   Bundesministerium f\"ur Forschung und Technologie under contract
   06 TM 732 and by the CEC science project SC1--CT91--0729.}}
\vskip1truecm
\centerline{\sc Andrzej J. Buras}
\bigskip
\centerline{\sl Technische Universit\"at M\"unchen, Physik Department}
\centerline{\sl D-85748 Garching, Germany}
\vskip0.6truecm
\centerline{\sl Max-Planck-Institut f\"ur Physik}
\centerline{\sl  -- Werner-Heisenberg-Institut --}
\centerline{\sl F\"ohringer Ring 6, D-80805 M\"unchen, Germany}

\vskip1truecm
\centerline{\bf Abstract}
We review the present status of CP violation in the standard model.
Subsequently we make an excursion in the future in order to see what
we could expect in this field in this and the next decade.
We present various strategies for the determination of the CKM parameters
and divide the decays into four classes with respect to theoretical
uncertainties.
 We emphasize that
the definitive tests of the Kobayashi-Maskawa picture of CP violation will
come
through a {\it simultaneous} study of CP asymmetries in $B_{d,s}^o$ decays,
the rare decays
$K^+ \to \pi^+\nu \bar\nu$ and $K_L \to \pi^o\nu\bar\nu$, and $x_d/x_s$.
We illustrate how the measurements of the CP asymmetries in
$B^0_{d,s}$-decays together with a measurement of
$Br(K_L\to \pi^\circ\nu\bar\nu)$ or $Br(\kpnn)$
and the known value of $\mid V_{us}\mid $ can determine {\it all} elements
of the Cabibbo-Kobayashi-Maskawa matrix essentially without any hadronic
uncertainties.

\newpage
\setcounter{page}{1}

\centerline{\Large\bf  CP Violation: Present and Future
   \footnote[1]{\noindent
   Supported by the German
   Bundesministerium f\"ur Forschung und Technologie under contract
   06 TM 732 and by the CEC science project SC1--CT91--0729.}}
\vskip1truecm
\centerline{\sc Andrzej J. Buras}
\bigskip
\centerline{\sl Technische Universit\"at M\"unchen, Physik Department}
\centerline{\sl D-85748 Garching, Germany}
\vskip0.6truecm
\centerline{\sl Max-Planck-Institut f\"ur Physik}
\centerline{\sl  -- Werner-Heisenberg-Institut --}
\centerline{\sl F\"ohringer Ring 6, D-80805 M\"unchen, Germany}
\vskip1truecm
\centerline{\bf Abstract}
{\small
We review the present status of CP violation in the standard model.
Subsequently we make an excursion in the future in order to see what
we could expect in this field in this and the next decade.
We present various strategies for the determination of the CKM parameters
and divide the decays into four classes with respect to theoretical
uncertainties.
 We emphasize that
the definitive tests of the Kobayashi-Maskawa picture of CP violation will
come
through a {\it simultaneous} study of CP asymmetries in $B_{d,s}^o$ decays,
the rare decays
$K^+ \to \pi^+\nu \bar\nu$ and $K_L \to \pi^o\nu\bar\nu$, and $x_d/x_s$.
We illustrate how the measurements of the CP asymmetries in
$B^0_{d,s}$-decays together with a measurement of
$Br(K_L\to \pi^\circ\nu\bar\nu)$ or $Br(\kpnn)$
and the known value of $\mid V_{us}\mid $ can determine {\it all} elements
of the Cabibbo-Kobayashi-Maskawa matrix essentially without any hadronic
uncertainties.
}


\section{Setting the Scene}
\subsection{The Cabibbo-Kobayashi-Maskawa Matrix}
In the Standard Model with three fermion generations, CP violation arises
from a single phase in the unitary $3\times 3$ Cabibbo-Kobayashi-Maskawa
matrix \cite{CAB,KM} which parametrizes the charged current interactions of
 quarks:
\begin{equation}\label{1j}
J^{cc}_{\mu}=(\bar u,\bar c,\bar t)_L\gamma_{\mu}
\left(\begin{array}{ccc}
V_{ud}&V_{us}&V_{ub}\\
V_{cd}&V_{cs}&V_{cb}\\
V_{td}&V_{ts}&V_{tb}
\end{array}\right)
\left(\begin{array}{c}
d \\ s \\ b
\end{array}\right)_L
\end{equation}
Following Wolfenstein \cite{WO}, it is useful but not necessary
to expand each
element of the CKM matrix as a power series in the small parameter
$\lambda=\mid V_{us}\mid=0.22 $:
\begin{equation}\label{2.75}
V_{CKM}=
\left(\begin{array}{ccc}
1-{\lambda^2\over 2}&\lambda&A\lambda^3(\varrho-i\eta)\\ -\lambda&
1-{\lambda^2\over 2}&A\lambda^2\\ A\lambda^3(1-\varrho-i\eta)&-A\lambda^2&
1\end{array}\right)
+O(\lambda^4)
\end{equation}
Because of the
smallness of $\lambda$ and the fact that for each element
the expansion parameter is actually
$\lambda^2$, it is sufficient to keep only the first few terms
in this expansion.
Following \cite{BLO} we will define the parameters
$(\lambda, A, \varrho, \eta)$ through
\b\l{wop}
s_{12}\equiv\lambda \qquad s_{23}\equiv A \lambda^2 \qquad
s_{13} e^{-i\delta}\equiv A \lambda^3 (\varrho-i \eta)      \e
where $s_{ij}$ and $\delta$ enter the standard parametrization \cite{PDG}
 of the CKM
matrix. This specifies the higher orders terms in (\ref{2.75}).

{}From tree level B decays sensitive to $V_{cb}$ and $V_{ub}$, the parameters
$A$, $\varrho$ and $\eta$ are constrained as follows \cite{STONE}:
\begin{equation}\label{2}
\lambda^2 A=\mid V_{cb} \mid=0.038\pm0.004
\end{equation}
\begin{equation}\label{2.94}
R_b \equiv  \sqrt{\bar\varrho^2 +\bar\eta^2}
= (1-\frac{\lambda^2}{2})\frac{1}{\lambda}
\left| \frac{V_{ub}}{V_{cb}} \right|=0.36\pm0.09
\end{equation}
where we have introduced \cite{BLO}
\begin{equation}\label{3}
\bar\varrho=\varrho (1-\frac{\lambda^2}{2})
\qquad
\bar\eta=\eta (1-\frac{\lambda^2}{2}).
\end{equation}
In order to determine $\varrho$ and $\mid\eta\mid$ we still need the value of
\begin{equation}\label{2.95}
R_t \equiv \sqrt{(1-\bar\varrho)^2 +\bar\eta^2}
=\frac{1}{\lambda} \left| \frac{V_{td}}{V_{cb}} \right|
\end{equation}
which is governed by $\mid V_{td}\mid$. From (\ref{2.94}) and (\ref{2.95}) we
have
$1-R_b\leq R_t \leq 1+R_b$
and unless $R_t=1\pm R_b$,
one finds $\eta\not=0$, which implies CP
violation in the standard model.

We observe that within the standard model the measurements of four CP
{\it conserving } decays sensitive to $\mid V_{us}\mid$, $\mid V_{ub}\mid$,
$\mid V_{cb}\mid $ and $\mid V_{td}\mid$ can tell us whether CP violation
is predicted in the standard model. This is a very remarkable property of
the Kobayashi-Maskawa picture of CP violation: quark mixing and CP violation
are closely related to each other. For this reason it is mandatory to discuss
here also the most important CP conserving decays.

All this can be shown transparently in the $(\bar\rho,\bar\eta)$ plane.
Starting with the unitarity relation
\begin{equation}\label{2.87h}
V_{ud}V_{ub}^* + V_{cd}V_{cb}^* + V_{td}V_{tb}^* =0,
\end{equation}
rescaling it by $\mid V_{cd}V_{cb}^\ast\mid=A \lambda^3$ and depicting
the result in the complex $(\bar\rho,\bar\eta)$ plane, one finds the
unitarity triangle of fig. 1. The lenghts CB, CA and BA are equal to 1,
$R_b$ and $R_t$ respectively. We observe that beyond the leading order
in $\lambda$ the point A does not correspond to $(\varrho,~\eta)$ but
to $(\bar\varrho,~\bar\eta)$. Clearly within 3\% accuracy
 $(\bar\varrho,~\bar\eta)=(\varrho,~\eta)$. In the distant future this
difference may matter however.

\vspace{6.0cm}
\centerline{Fig. 1}
\vspace{0.2cm}

The triangle in fig. 1 is one of the important targets of the contemporary
particle physics. Together with $\mid V_{us}\mid$ and $\mid V_{cb}\mid$
it summarizes the structure of the CKM matrix. In particular the area of
the unrescaled triangle gives a measure of CP violation in the
standard model
\cite{JAR}:
\begin{equation}\label{555}
\mid J_{CP}\mid=
2\cdot ({\rm Area~ of~} \Delta)=
\mid V_{ud}\mid\mid V_{us}\mid\mid V_{ub}\mid\mid V_{cb}\mid \sin\delta=
A^2\lambda^6\bar\eta=0(10^{-5}).
\end{equation}
This formula shows another important feature of the KM picture of CP
violation: the smallness of CP violation in the standard model is not
necessarily related to the smallness of $\eta$ but to the fact that in this
model the size of CP violating effects is given by products of
small mixing parameters.

Since the top quark mass is an important parameter in the field of
CP violation, we have to specify what we mean by $m_t$. Here in accordance
with various QCD calculations quoted below, we will use
$m_t\equiv \overline{m_t}(m_t)$: the current top
quark mass at the scale $m_t$. The physical top quark mass ($m_t^{phys}$)
defined as
the pole of the renormalized propagator is by about $7~GeV$ higher than
$m_t$.

Finally it should be stated that a large part in the errors quoted in
(\ref{2}) and (\ref{2.94}) results from theoretical uncertainties.
Consequently even if the data from CLEO improves in the
future, it is difficult to imagine at present that in the tree level B-decays
a better accuracy than $\Delta\vcb=\pm 2\cdot 10^{-3}$ and
$\Delta\vub=\pm 0.01$ ($\Delta R_b=\pm 0.04$) could be achieved \cite{BALL}.
We will see below that the loop induced decays governed by short distance
physics can in principle offer a more accurate determination of $\vcb$ and
$\vub$.
\subsection{Loop induced Decays and Transitions}
Using  (\ref{2}), (\ref{2.94}) and (\ref{2.95})
 we find
$\mid V_{td}\mid\leq A\lambda^3 (1+R_b)\leq 13.4\cdot 10^{-3}$
and the branching ratio $Br(t\to d)\leq 10^{-3}$. Consequently it will be
very difficult to measure $\vtd$ in tree level top quark decays. In order
to find $\vtd$ we have to measure loop induced decays and transitions
governed by penguin and box diagrams with internal top quark exchanges.

In the {\it K-meson} system the top favourites are: the indirect
($\varepsilon_K$)
and the direct ($\varepsilon'$) CP violating contributions to $K\to\pi\pi$,
the rare decays $K_L\to\pi^\circ e^+e^-,~K_L\to\mu\bar\mu,
{}~K^+\to\pi^+\nu\bar\nu,~K_L\to\pi^\circ\nu\bar\nu$ and the parity violating
asymmetry $\Delta_{LR}$ in $K^+\to\pi^+\mu^+\mu^-$.

In the {\it B-meson} system the corresponding favourites are:
$B^\circ_d-\bar B^\circ_d$ ($B^\circ_s-\bar B^\circ_s$) mixing described
by the parameter $x_d$ ($x_s$) and the rare decays
$B\to\mu\bar\mu,~B\to X_{d,s}\nu\bar\nu$, $B\to X_{d,s}\gamma$ and
 $B \to X_{d,s} e^+e^-$.

Furthermore a very special role is played by CP-asymmetries in the decays
$B_{d,s}^\circ\to f$ where $f$ is a CP eigenstate. Some of these
asymmetries determine the angles in fig.1 $(\alpha,~\beta,~\gamma)$ without
any theoretical uncertainties \cite{NQ}. Consequently their measurements
 will have
important impact on the search of the unitarity triangle ($\Delta$) and
indirectly on $\mid V_{td}\mid$. We will return to CP asymmetries in
sections 5-7.

{}From this long list only $\varepsilon_K$ and $x_d$ are useful for $\Delta$
at present but in 15 years from now the picture of $\Delta$ might well look
like the one shown in fig.2.

The general structure of theoretical expressions for the relevant decay
amplitudes
is given in a simplified form  roughly as follows:
\begin{equation}\label{6}
{\rm A(Decay)} =B V_{CKM}\eta_{QCD}F(m_t)+({\rm Charm~Contributions})+
({\rm LD~Contributions})
\end{equation}
Here $V_{CKM}$ represents a given product of the CKM elements we want to
determine. $F(m_t)$ results from the evaluation of loop diagrams with top
exchanges and $\eta_{QCD}$ summarizes short distance QCD corrections to
a given decay. By now these corrections are known essentially for all
decays listed above at the leading and next-to-leading order in the
renormalization group improved perturbation theory. Next $B$ stands for
a non-perturbative factor related to the relevant hadronic matrix element of
the contributing four fermion operator: the main theoretical uncertainty in
the whole
enterprise. In semi-leptonic decays such as $K\to \pi\nu\bar\nu$,
the non-perturbative $B$-factors can
fortunately be determined from leading tree level decays such as
$K^+\to\pi^\circ e^+\nu$ reducing or removing the
non-perturbative uncertainty. In non-leptonic decays and in
$B^\circ-\bar B^\circ$ mixing this is generally not possible and we have
to rely on existing non-perturbative methods. The additional terms
in (\ref{6}) include internal charm contributions and sometimes
unwanted
long distance contributions as in the case of $K_L\to\mu\bar\mu$. In
B-decays the internal top contributions are essentially the whole story.
In K-decays
the internal charm contributions can sometimes be also important as in
the case of $K^+\to\pi^+\nu\bar\nu$. Finally in more complicated decays,
in particular in $\varepsilon'$, one finds linear combinations of different
$m_t$-dependent functions. Moreover due to the appearance of several
contributing operators a set of different B-factors can be present.

\vspace{11.5cm}
\centerline{Fig. 2}

\subsection{Classification}
Let us group the various decays and quantities in four different classes
with respect to hadronic uncertainties present in them.
\begin{itemize}
\item Class I (Essentially no hadronic uncertainties):\\
 $K_L\to \pi^\circ\nu\bar\nu$ and some
 CP asymmetries in neutral B decays to CP
eigenstates which give $\sin 2\alpha,~\sin2\beta,~\sin2\gamma$.
\item Class II (Small theoretical uncertainties related to
$\Lambda_{\overline{MS}}$, $m_c$, the renormalization scale
$\mu$ or $SU(3)$-breaking effects.):\\
$K^+\to\pi^+\nu\bar\nu$, the parity violating asymmetry
$\Delta_{LR}$, $x_d/x_s$ and
 $B\to X_{d,s}\nu\bar\nu$.
\item Class III (Hadronic uncertainties are present but will probably
be reduced considerably in the next five years):\\
$K_L\to\pi^\circ e^+e^-,~\varepsilon_K,~x_{d,s},
{}~B_{d,s}\to\mu\bar\mu,~B\to X_{d,s}\gamma$, $ B\to X_{d,s} e^+ e^-$.
\item Class IV (Large hadronic uncertainties which can only be removed
if some dramatic improvements in non-perturbative techniques will take
place):\\ $\varepsilon',~K_L\to\mu\bar\mu$,~CP asymmetries in $B^{\pm}$ and
hyperon decays.
\end{itemize}
It should be stressed that all these decays are very interesting.
In particular, in addition to $\klpnn$ in class I, also
 ${\rm Re}(\varepsilon'/\varepsilon)$ and
$B^+$ decays in class IV may give first signals of the direct CP violation.
However only
the decays of class I and II, when measured, allow a reliable
 determination of
CKM parameters unless considerable
improvements in non-perturbative techniques will
be made. The $B$-factors and long distance contributions in class III
are easier to calculate than for quantities in class IV.
\section{Strategy}
During the coming fifteen years we will certainly witness a dramatic
improvement in the determination of the CKM-parameters analogous
to, although not as precise as, the determination of the parameters
in the gauge boson sector which took place during the last years.
Let us recall that the relevant {\it independent} parameters in the
electroweak precision studies are:
\begin{equation}\label{EW}
G_F, \quad \alpha_{QED}(m_e), \quad M_Z, \quad m_t, \quad  m_H
\end{equation}
with $\alpha_{QCD}$ or $\Lambda_{\overline{MS}}$ playing sometimes
some role. In the field of quark mixing and CP violation the
relevant parameters are
\begin{equation}
\lambda, \quad A, \quad \varrho, \quad\eta, \quad m_t
\end{equation}
with $\Lambda_{\overline{MS}}$ and $m_c$ playing often
sizable roles. Moreover as stated above, non-perturbative
effects in class III and IV decays play a very important role.
On the other hand due to the natural flavour conservation in neutral
current processes and the small couplings of the neutral higgs to $s$
and $b$ quarks, the impact of $m_H$ on this field can be fully
neglected.

There is of course most probably and hopefully some new physics
beyond the standard model. This physics introduces generally
new parameters and makes
the full discussion considerably more involved. Moreover theoretical
calculations, in particular of the QCD corrections, for the extensions
of the standard model are often not at
 the level of existing calculations in
this model and consequently no precise discussion of various effects
related to new physics is possible at present.

Our strategy then will be to confine the presentation exclusively to the
standard model and to discuss several quantities simultaneously with
the hope that future precise measurements will display some inconsistencies
which will signal a new physics beyond the standard model. Moreover we
will devote a large part of this review to decays of class I and II which
being essentially free from hadronic uncertainties are ideally suited for
the determination of the CKM parameters. We will
however also discuss some of the decays of class III and IV.
\section{Messages from the Indirect CP Violation}
The indirect CP violation in the K-system discovered in 1964 \cite{CRO}
and parametrized
by $\varepsilon_K$ is the only clear experimental signal of this important
phenomenon. The usual box diagram calculation together with the experimental
value $\varepsilon_K=2.26\cdot 10^{-3}\exp(i\pi/4)$ specifies a hyperbola in
the $(\bar\rho,\bar\eta)$ plane with $\bar\eta>0$ as shown in fig.2. The
position of this hyperbola depends on $m_t$, $V_{cb}$ and on
the non-perturbative parameter $B_K$. There are essentially four messages
here:
\begin{itemize}
\item For a given set $(m_t,~V_{cb},~\mid V_{ub}/V_{cb}\mid,~B_K)$
one determines two values of $\eta$ corresponding to two crossing points
of the $\epsilon_K$-hyperbola with the circle (\ref{2.94}).
Typically $ 0.20<\eta<0.45$ is found.
\item With decreasing $\mid V_{cb}\mid,~B_K$ and $m_t$, the $\epsilon_K$-
hyperbola moves away from the origin of the $(\bar\rho,\bar\eta)$ plane.
When the hyperbola and the circle (\ref{2.94}) only touch each other a lower
bound for $m_t$ follows \cite{AB}:
\begin{equation}\label{115}
(m_t)_{min} = M_W \left[ \frac{1}{2 A^2} \left(\frac{1}{A^2 B_K R_b}
 - 1.2 \right) \right]^{0.658}
\end{equation}
For $V_{cb}=0.040,~\mid V_{ub}/V_{cb}\mid=0.08$ and $B_K=0.75$ one has
$(m_t)_{min}=170~GeV.$
\item For a given $m_t$ a lower bound on $\mid V_{cb}\mid,
 ~\mid V_{ub}/V_{cb}\mid$ and $B_K$ can be found. For instance \cite{BLO}
\begin{equation}\label{116}
(B_K)_{min} = \left[ A^2 R_b \left( 2 x_t^{0.76} A^2 + 1.2 \right)
\right]^{-1}
;\qquad x_t=\frac{m_t^2}{M^2_W}
\end{equation}
\item The CDF value for $m_t$ ($m_t^{phys}=174\pm 16~GeV$) \cite{CDF}
 together with
(\ref{115}) and (\ref{116}) imply that the observed indirect CP
violation can be accomodated in the standard model provided
$\vcb$, $\vub$ and $B_K$ are not too small.
For instance with $m_t<180~GeV$, $\mid V_{ub}/V_{cb}\mid<0.09$ and
$\mid V_{cb}\mid<0.040$ only values $B_K>0.62$ are allowed. Such
values are found in the lattice ($B_K=0.83\pm0.03$) \cite{SHARP} and
the 1/N approach ($B_K=0.7\pm 0.1$) \cite{BBG}.
\end{itemize}
To summarize: the presently measured values of $\vub$, $\vcb$ and
$m_t$ together with the non-perturbative calculations of $B_K$ imply
that that the KM picture of CP violation is consistent with the
bound in (\ref{115}). In view of large uncertainties
in the
four input parameters in question this first test of the KM picture
is however by no means conclusive.
\section{$\varepsilon'/\varepsilon,~K_L\to\pi^\circ e^+e^-,~
K_L\to\pi^\circ\nu\bar\nu,~K^+\to\pi^+\nu\bar\nu$}
Let me now discuss four stars in the field of K-decays. The first
three deal with searches of direct CP violation. The last one is
CP conserving but plays an important role in the determination of the
unitarity triangle. The fifth star, the parity violating asymmetry
in $K^+\to\pi^+\mu^+\mu^-$ ($\Delta_{LR}$) \cite{SWW}
is discussed elsewhere \cite{BB94A}.
\subsection{$\varepsilon'/\varepsilon$}
Re($\varepsilon'/\varepsilon$) measures the ratio of direct to indirect
CP violation in $K\to\pi\pi$ decays. In the standard model
$\varepsilon'/\varepsilon $ is governed by QCD penguins and electroweak (EW)
penguins \cite{GIL0}. In spite of being suppressed by $\alpha/\alpha_s$
relative to QCD penguin contributions, the electroweak penguin contributions
have to be included because of the additional enhancement factor
$ReA_o/ReA_2=22$ relative to QCD penguins. Moreover with increasing $m_t$
the EW-penguins become increasingly important \cite{FLYNN,BBH} and entering
$\varepsilon'/\varepsilon$ with the opposite sign to QCD-penguins suppress
this ratio for large $m_t$. For $m_t\approx 200~GeV$ the ratio can even
be zero \cite{BBH}.
 The short distance QCD corrections to
$\varepsilon'/\varepsilon$ are known at the NLO level \cite{BJLW,ROMA}.
Unfortunately
$\varepsilon'/\varepsilon$ is plagued with uncertainties related to
non-perturbative B-factors which multiply $m_t$ dependent functions in a
formula like (\ref{6}). Several of these B-factors can be extracted from
leading
CP-conserving $K\to\pi\pi$ decays \cite{BJLW}. Two important B-factors
($B_6=$ the dominant QCD penguin and $B_8=$ the dominant electroweak penguin)
cannot be determined this way and one has to use lattice or $1/N$ methods
to predict Re($\varepsilon'/\varepsilon$). An analytic formula for
Re($\varepsilon'/\varepsilon$) as a function of
$m_t,~\Lambda_{\overline{MS}},~B_6,~B_8,~m_s$ and $V_{CKM}$ can be found
in \cite{BLAU}. A very simplified version of this formula is given as follows
\begin{equation}\label{7e}
{\rm Re}(\frac{\varepsilon'}{\varepsilon})=12\cdot 10^{-4}\left [
\frac{\eta\lambda^5 A^2}{1.7\cdot 10^{-4}}\right ]
\left [\frac{150~MeV}{m_s(m_c)} \right ]^2 [B_6-Z(x_t)B_8]
\end{equation}
where $Z(x_t)$ is given in (\ref{11k}).
For $m_t=170\pm10~GeV$ and using $\varepsilon_K$-analysis to determine
$\eta$ one finds \cite{BJLW,ROMA}
\begin{equation}\label{8}
2\cdot 10^{-4} \leq \frac{\varepsilon'}{\varepsilon}\leq 1\cdot 10^{-3}
\end{equation}
if $B_6\approx B_8\approx 1$ (lattice, 1/N expansion) are used.
For $B_6\approx 2$
and $B_8\approx 1$ as advocated in \cite{DORT},
Re($\varepsilon'/\varepsilon$) increases to $(15\pm5)\cdot 10^{-4}$.

The experimental situation on Re($\varepsilon'/\varepsilon$) is unclear
at present.
 While
the result of NA31 collaboration at CERN with Re$(\epsilon'/\epsilon)
= (23 \pm 7)\cdot 10^{-4}$ \cite{WAGNER} clearly indicates
direct CP violation, the value of E731 at Fermilab,
Re$(\epsilon'/\epsilon) = (7.4 \pm 5.9)\cdot 10^{-4}$
\cite{GIBBONS} is compatible with superweak theories \cite{WO1} in which
$\epsilon'/\epsilon = 0$.
Both results are in the ball park of the theoretical estimates although
the NA31 result appears a bit high compared to the range given in
(\ref{8}).

 Hopefully, in about five years the
experimental situation concerning $\epsilon'/\epsilon$ will be
clarified through the improved measurements by the two collaborations
at the $10^{-4}$ level and by experiments at the $\Phi$ factory in
Frascati.
One should also hope that the theoretical situation of
$\varepsilon'/\varepsilon$ will improve by then to confront the new data.
\subsection{$K_L\to\pi^o e^+e^-$}
       Whereas in $K \to \pi \pi$ decays the CP violating
contribution is a tiny part of the full amplitude and the direct CP
violation is expected to be at least by three orders of magnitude
smaller than the indirect CP violation, the corresponding hierarchies
are very different for the rare decay $K_L\to\pi^o e^+e^-$ .
At lowest order in
electroweak interactions (single photon, single Z-boson or double
W-boson exchange), this decay takes place only if CP symmetry is
violated \cite{GIL1}.
Moreover, the direct CP violating contribution is predicted to be
larger than
the indirect one. The CP conserving contribution to the amplitude
comes from a two photon exchange, which
although higher order in $\alpha$ could in principle be sizable.  The studies
of
the last years \cite{PICH} indicate however that the
CP conserving part is significantly smaller than the direct CP
violating contribution.

The size of the indirect CP violating contribution will be
known once the CP conserving decay $K_S \to \pi^0 e^+ e^-$ has been
measured \cite{BARR}. On the other hand the direct CP violating
contribution can
be fully calculated as a function of $m_t$, CKM parameters and the
QCD coupling constant $\alpha_s$. There are practically no theoretical
uncertainties related to hadronic matrix elements in this part,
because the latter can be extracted from the well-measured decay $K^+
\to \pi^0 e^+ \nu$.
The next-to-leading QCD corrections to the direct CP violating contribution
have been recently calculated \cite{BLMM}
reducing certain ambiguities present in
leading order analyses \cite{GIL2} and enhancing the leading order results by
roughly $25\%$. The final result is given by
\begin{equation}\label{9a}
Br(K_L \to \pi^0 e^+ e^-)_{dir} = 4.16\cdot (Im\lambda_t)^2
(y_{7A}^2 + y_{7V}^2)
\end{equation}
where
$Im \lambda_t = Im (V_{td} V^*_{ts})$
and
\begin{eqnarray}
\label{pbe7v}
y_{7V} & = & \f{\alpha}{2\pi \sin^2 \theta_W} \left( P_0 +
Y(x_t) - 4 \sin^2 \theta_W \; Z(x_t) \right), \\
\label{pbe7a}
y_{7A} & = & - \f{\alpha}{2\pi \sin^2 \theta_W} Y(x_t).
\end{eqnarray}

Here, to a very good approximation for $140~GeV\leq m_t \leq 230~GeV$,
\begin{equation}\label{11k}
Y(x_t) = 0.315\cdot x_t^{0.78}, \hspace{2cm} Z(x_t) = 0.175\cdot x_t^{0.93}.
\end{equation}
Next $P_o=0.70\pm0.02$ as found in \cite{BLMM}.
For $m_t=170\pm 10~GeV$ one finds
\begin{equation}\label{12}
Br(K_L\to\pi^0 e^+ e^-)_{dir}=(5.\pm 2.)\cdot 10^{-12}
\end{equation}
where the error comes dominantly from the uncertainties in the CKM
parameters.
This should be compared with the present estimates of the other two
contributions:  $Br(K_L\to\pi^o e^+e^-)_{indir}\leq 1.6\cdot 10^{-12}$
and $Br(K_L\to\pi^o e^+e^-)_{cons}\approx(0.3-1.8)\cdot 10^{-12}$ for
the indirect
CP violating and the CP conserving contributions respectively \cite{PICH}.
Thus direct
CP violation is expected to dominate this decay.

The present experimental
bounds
\begin{equation}
Br(K_L\to\pi^0 e^+ e^-) \leq\left\{ \begin{array}{ll}
4.3 \cdot 10^{-9} & \cite{harris} \\
5.5 \cdot 10^{-9} & \cite{ohl} \end{array} \right.
\end{equation}
are still by three orders of magnitude away from the theoretical
expectations in the Standard Model. Yet the prospects of getting the
required sensitivity of order $10^{-11}$--$10^{-12}$ in six years are
encouraging
\cite{CPRARE}.
\subsection{$K_L\to\pi^o\nu\bar\nu$ and $K^+\to\pi^+\nu\bar\nu$}
$K_L\to\pi^o\nu\bar\nu$ and $K^+\to\pi^+\nu\bar\nu$ are the theoretically
cleanest decays in the field of rare K-decays.
$K_L\to\pi^o\nu\bar\nu$ is dominated by short distance loop diagrams
involving
the top quark and proceeds almost entirely through direct CP violation
\cite{LI}. $K^+\to\pi^+\nu\bar\nu$ is CP conserving and receives
contributions from
both internal top and charm exchanges.
The last year calculations
\cite{BB2,BB3} of next-to-leading QCD corrections to these decays
considerably reduced the theoretical uncertainty
due to the choice of the renormalization scales present in the
leading order expressions \cite{DDG}. Since the relevant hadronic matrix
elements of the weak current $\bar {s} \gamma_{\mu} (1- \gamma _{5})d $
can be measured in the leading
decay $K^+ \rightarrow \pi^0 e^+ \nu$, the resulting theoretical
expressions for Br( $K_L\to\pi^o\nu\bar\nu$) and Br($K^+\to\pi^+\nu\bar\nu$)
  are
only functions of the CKM parameters, the QCD scale
 $\Lambda \overline{_{MS}}$
 and the
quark masses $m_t$ and $m_c$.
The long distance contributions to
$K^+ \rightarrow \pi^+ \nu \bar{\nu}$ have been
considered in \cite{RS} and found to be very small: two to three
orders of magnitude smaller than the short distance contribution
at the level of the branching ratio.
The long distance contributions to $K_L\to\pi^o\nu\bar\nu$ are negligible
as well.

The explicit expressions for $Br(\kpnn)$ and $Br(\klpnn)$ are given as
 follows
\b\l{bkpn}
Br(\kpn)=4.64\cdot 10^{-11}
\cdot\left[\left({\imlt\o\lambda^5}X(x_t)\right)^2+
\left({\relc\o\lambda}P_0(K^+)+{\relt\o\lambda^5}X(x_t)\right)^2
\right]            \e
\b\l{bklpn}
Br(K_L\to\pi^0\nu\bar\nu)=1.94\cdot 10^{-10}
\cdot\left({\imlt\o\lambda^5}X(x_t)  \right)^2    \e
$\lambda_c$
is essentially real and
$X(x_t)$ is given to an excellent accuracy by
\begin{equation}\label{xt}
X(x_t) = 0.65\cdot x_t^{0.575}
\end{equation}
where the NLO correction calculated in \cite{BB2} is included if
$m_t\equiv\bar m_t(m_t)$.
 Next $P_0(K^+)=0.40\pm0.09$ \cite{BB3,BB4} is a function of $m_c$ and
 $\Lambda_{\overline{MS}}$ and includes the residual uncertainty
due to the renormalization scale $\mu$. The absence of $P_0$ in
(\ref{bklpn}) makes $\klpnn$ theoretically even cleaner than $\kpnn$.

It has been pointed out in \cite{BH} that
measurements of $Br(\kpn)$ and $Br(\klpn)$ could determine the
unitarity triangle completely provided $m_t$ and $V_{cb}$ are known.
Generalizing this analysis to include non-leading terms in $\lambda$
one finds to a very good accuracy \cite{BB4}
($\sigma=(1-\lambda^2/2)^{-2}$):
\begin{equation}\label{18}
\bar\varrho=1+{P_0(K^+)-\sqrt{\sigma(B_+-B_L)}\o A^2 X(x_t)}\qquad
\bar\eta={\sqrt{B_L}\o\sqrt{\sigma} A^2 X(x_t)}
\end{equation}
where we have introduced the "reduced" branching ratios
\begin{equation}\label{19}
B_+={Br(\kpn)\o 4.64\cdot 10^{-11}}\qquad
B_L={Br(\klpn)\o 1.94\cdot 10^{-10}}
\end{equation}

It follows from (\ref{18}) that
\begin{equation}\label{21}
r_s(B_+, B_L)={1-\bar\varrho\o\bar\eta}=\cot\beta =
\sqrt{\sigma}{\sqrt{\sigma(B_+-B_L)}-P_0(K^+)\o\sqrt{B_L}}
\end{equation}
so that
\begin{equation}\label{22}
\sin 2\beta=\frac{2 r_s(B_+, B_L)}{1+r^2_s(B_+, B_L)}
\end{equation}
does not depend on $m_t$ and $V_{cb}$. An exact treatment of the CKM matrix
confirms this finding to high accuracy.
Consequently $\kpnn$ and $\klpnn$ offer a
clean
determination of $\sin 2\beta$ which can be confronted with
the one possible in $B^0\to\psi K_S$ discussed below.
Combining these two ways of
determining $\sin 2\beta$ one finds an interesting relation
 between rare K decays and B physics
\begin{equation}\label{23}
{2 r_s(B_+,B_L)\o 1+r^2_s(B_+,B_L)}=-A_{CP}(\psi K_S){1+x^2_d\o x_d}
\end{equation}
which must be
satisfied in the standard model. Any deviation from this relation
would signal new physics. A numerical analysis of
(\ref{bkpn}), (\ref{bklpn})
and (\ref{22}) will be given below.
\section{CP Asymmetries in B-Decays and $x_d/x_s$}
\subsection{CP-Asymmetries in $B^o$-Decays}
The CP-asymmetry in the decay $B_d^\circ \rightarrow \psi K_S$ allows
 in the standard model
a direct measurement of the angle $\beta$ in the unitarity triangle
without any theoretical uncertainties
\cite{NQ}. Similarly the decay
$B_d^\circ \rightarrow \pi^+ \pi^-$ gives the angle $\alpha$, although
 in this case strategies involving
other channels are necessary in order to remove hadronic
uncertainties related to penguin contributions
\cite{CPASYM}.
The determination of the angle~$\gamma$ from CP asymmetries in neutral
B-decays is more difficult but not impossible
\cite{RF}. Also charged B decays could be useful in
this respect \cite{Wyler}.
We have for instance
\begin{equation}\label{113c}
 A_{CP}(\psi K_S)=-\sin(2\beta) \frac{x_d}{1+x_d^2},~~~~~~~~~~
   A_{CP}(\pi^+\pi^-)=-\sin(2\alpha) \frac{x_d}{1+x_d^2}
\end{equation}
where we have neglected QCD penguins in $A_{CP}(\pi^+\pi^-)$.
Since in the triangle of fig.1 one side is known, it suffices to measure
two angles to determine the triangle completely. We will investigate the
impact of the future measurements of $\sin 2\alpha$ and $\sin 2\beta$ below.
 $\sin(2\phi_i)$ can be expressed
in terms of $(\bar\varrho,\bar\eta)$ as follows \cite{BLO}
\begin{equation}\label{1}
\sin(2\alpha)=\frac{2\bar\eta(\bar\eta^2+\bar\varrho^2-\bar\varrho)}
  {(\bar\varrho^2+\bar\eta^2)((1-\bar\varrho)^2
  +\bar\eta^2)}
\qquad
\sin(2\beta)=\frac{2\bar\eta(1-\bar\varrho)}{(1-\bar\varrho)^2 + \bar\eta^2}
\end{equation}
We will see below that the asymmetry $A_{CP}(\psi K_S)$ could be as high
as --0.4. This is not in contradiction with (\ref{555}) because
 the corresponding
branching ratio for this decay is $O(10^{-4})$. This possibility of
observing large CP asymmetries in B-decays makes them particulary useful
for the tests of the KM picture.
\subsection{$B^o-\bar B^o$ Mixing}
Measurement of $B^o_d-\bar B^o_d$ mixing parametrized by $x_d$ allows to
determine $R_t$:
\begin{equation}\label{106}
 R_t = 1.63 \cdot \frac{R_0}{\sqrt{ S(x_t)}}\qquad
 S(x_t) = 0.784 \cdot x_t^{0.76}
\end{equation}
where
\begin{equation}\label{107}
R_0 \equiv \sqrt{ \frac{x_d}{0.72}} \left[ \frac{200 MeV}{F_{B_d}
 \sqrt{B_{B_d}}}
\right] \left[ \frac{0.038}{\kappa} \right] \sqrt{ \frac{0.55}{\eta_B}}
\qquad
\kappa \equiv \mid V_{cb} \mid \left[\frac{\tau_B}{1.5\,ps}\right]^{0.5}
\end{equation}
with $\tau_B$ being the B-meson life-time.
$\eta_B=0.55$ is the QCD factor
\cite{BJW}.
$F_{B_d}$ is the B-meson decay constant and $B_{B_d}$
denotes a non-perturbative
parameter analogous to $B_K$. The values of $x_d$, $F_{B_d} \sqrt{ B_{B_d}}$
and $|V_{cb}|$  will be specified below.

The accuracy of the determination of $R_t$ can be considerably improved
by
measuring simultaneously the $B^o_s-\bar B^o_s$ mixing described by $x_s$.
We have
\begin{equation}\label{107b}
R_t = \frac{1}{\sqrt{R_{ds}}} \sqrt{\frac{x_d}{x_s}} \frac{1}{\lambda}
\sqrt{1-\lambda^2(1-2 \varrho)}\qquad
R_{ds} = \frac{\tau_{B_d}}{\tau_{B_s}} \cdot \frac{m_{B_d}}{m_{B_s}}
\left[ \frac{F_{B_d} \sqrt{B_{B_d}}}{F_{B_s} \sqrt{B_{B_s}}} \right]^2
\end{equation}
Note that $m_t$ has been eliminated this way and $R_{ds}$ depends only on
SU(3)-flavour breaking effects which contain much smaller theoretical
uncertainties than the hadronic matrix elements in $x_d$ and $x_s$
separately. An estimate of such effects in
$F_{B_d} \sqrt{B_{B_d}}/F_{B_s} \sqrt{B_{B_s}}$ \cite{Grinstein} shows
that provided $x_d/x_s$ has been accurately measured a determination
of $R_t$ within $\pm 10\%$ should be possible. We will soon see that
a much more accurate determination of $R_t$ can be achieved by
measuring CP asymmetries in B-decays.
\section{Future Visions}
Here I would like to report on the results of recent studies presented
in detail in refs. \cite{BLO,BB4,AB94}. After showing the present
picture of the unitarity triangle corresponding to the range of parameters
given in (\ref{200}) we will
investigate what the future could bring us in this field.
Several lines of attack will be presented in this section culminating
with a precise determination of all CKM parameters in section 7.
\subsection{$\varepsilon_K,~B^o_d-\bar B^o_d,~\mid V_{ub}/V_{cb}\mid,
{}~\mid V_{cb} \mid $}
Here we just use the four quantities listed above
anticipating improved determinations of $m_t,~\mid V_{ub}/V_{cb}\mid,~
B_K,~F_B\sqrt{B_B}$ and $x_d$ in the next ten years (ranges II and III).
The measurements by CLEO and at LEP will play important roles here.
In view of our remarks in section 1.1, the range III assumes also
improvements in the theory.
We consider the following ranges \cite{BLO}:
\newpage

\underline{Range I}

\begin{equation}\label{200}
\begin{array}{rclrcl}
\left| V_{cb} \right| & = &  0.038 \pm 0.004 &
\mid V_{ub}/V_{cb} \mid & = & 0.08 \pm 0.02 \\
B_K & = & 0.7 \pm 0.2 & \sqrt{B_{B_d}} F_{B_d} & = & (200 \pm 30)~MeV \\
x_d & = & 0.72 \pm 0.08 & m_t & = & (165 \pm 15)~GeV \\
\end{array}
\end{equation}

\underline{Range II}
\begin{equation}\label{210}
\begin{array}{rclrcl}
\left| V_{cb} \right| & = &  0.040 \pm 0.002 &
\mid V_{ub}/V_{cb} \mid& = & 0.08 \pm 0.01 \\
B_K & = & 0.75 \pm 0.07 & \sqrt{B_{B_d}} F_{B_d} & = & (185 \pm 15)~MeV \\
x_d & = & 0.72 \pm 0.04 & m_t & = & (170 \pm 7)~GeV \\
\end{array}
\end{equation}

\underline{Range III}
\begin{equation}\label{211}
\begin{array}{rclrcl}
\left| V_{cb} \right| & = &  0.040 \pm 0.001 &
\mid V_{ub}/V_{cb} \mid & = & 0.08 \pm 0.005 \\
B_K & = & 0.75 \pm 0.05 & \sqrt{B_{B_d}} F_{B_d} & = & (185 \pm 10)~MeV \\
x_d & = & 0.72 \pm 0.04 & m_t & = & (170 \pm 5)~GeV \\
\end{array}
\end{equation}

The resulting unitarity triangles for ranges I-III are shown in
in the left half of fig. 3.
For the range III one has the following expectations:
\begin{equation}\label{13}
\begin{array}{rclrcl}
\sin2\alpha & = &  0.50 \pm 0.49 &
\sin2\beta & = & 0.61 \pm 0.09 \\
\mid V_{td}\mid & = & (9.4\pm 1.0)\cdot 10^{-3}&
x_s & = & 12.9\pm 2.8 \\
Br(K^+\to \pi^+\nu\bar\nu) & = & (1.03\pm 0.15)\cdot 10^{-10} &
Br(K_L\to \pi^\circ\nu\bar\nu) & = & (2.7\pm 0.4)\cdot 10^{-11} \\
\end{array}
\end{equation}
and $\sin(2\gamma)=0.\pm 0.68$.
We should remark that for the ranges II and III,
the uncertainties in $Br(\kpnn)$ due to $m_c$, $\Lambda_{\overline{MS}}$
and $\mu$ have been omitted. They will be included in sections 6.2 and
7.

This exercise implies that if the accuracy of various parameters given
in (\ref{210}) and (\ref{211}) is achieved, the determination of
$\mid V_{td} \mid$ and the predictions for $\sin(2\beta)$, $Br(K^+
\to \pi^+ \nu \bar\nu)$  and  $Br(K_L\to \pi^\circ\nu \bar\nu)$
are quite accurate. A sizable uncertainty in
$x_s$ remains however.
Another important message from this analysis is the inability of a
precise determination of $\sin(2\alpha)$ and $\sin(2\gamma)$ on the
basis of $\varepsilon_{K}$, $B^o - \bar{B^o}$, $|V_{cb}|$ and
$|V_{ub}/V_{cb}|$ alone.
This analysis shows that even with
the improved values of the parameters in question as given in
(\ref{210}) and (\ref{211}) a precise determination of $\sin(2\alpha)$
and $\sin(2\gamma)$  this way should not be expected.

\subsection{$\sin (2\beta)$ from $K\to \pi\nu\bar\nu$}
The numerical analysis of (\ref{18})--(\ref{22}) shows \cite{BB4} that
provided
$B(\kpn)$ and $B(\klpn)$ are measured within $\pm 10\%$ accuracy,
$\Delta\sin 2\beta=\pm 0.11$ could be achieved this way.
With decreasing uncertainty in $\Lambda_{\overline{MS}}$ and $m_c$
this error could be reduced to
$\Delta\sin 2\beta < \pm 0.10$.
The determination of $\sin 2\alpha$ and
$\sin 2\gamma$ on the other hand is rather poor. However
respectable determinations of the Wolfenstein parameter
$\eta$ and of $\mid V_{td}\mid$ can be obtained.
Choosing
$Br(\kpn)=(1.0\pm 0.1)\cdot 10^{-10}$,
$Br(\klpn)=(2.5\pm 0.25)\cdot 10^{-11}$,
 $m_t=(170\pm 5)GeV$
and $\mid V_{cb}\mid =0.040\pm 0.001$
one finds \cite{BB4}
\begin{equation}\label{26}
\sin(2 \beta)=0.60\pm 0.11 \qquad
\eta=0.34\pm 0.05 \qquad
\mid V_{td}\mid=(9.3\pm 2.1)\cdot 10^{-10}
\end{equation}

\vspace{16.1cm}
\centerline{Fig. 3}

\subsection{CP Asymmetries in $B^o$-Decays}
Let us next investigate the impact of the measurements of $\sin(2\alpha)$
and $\sin(2\beta)$ on the determination of the unitarity triangle.
As a warming up let us consider the accuracies
$\Delta \sin(2\beta) = \pm 0.06$ and
$\Delta \sin(2\alpha) = \pm 0.10$
which could be achieved around the year 2000 \cite{BAB,AL}.

In the right half of fig. 3 \cite{BLO} we show the
impact of such measurements taking
\begin{equation}\label{220}
\sin(2 \beta) = \left\{ \begin{array}{r}
0.60 \pm 0.18 \qquad (\rm{a}) \\
0.60 \pm 0.06 \qquad (\rm{b})
\end{array}\right.
\qquad
\sin(2 \alpha) = \left\{ \begin{array}{rc}
-0.20 \pm 0.10 & (\rm{I})  \\
0.10 \pm 0.10  & (\rm{II}) \\
0.70 \pm 0.10  & (\rm{III})
\end{array}\right.
\end{equation}
We observe that the measurement of $\sin(2\alpha)$
in conjunction with $\sin(2\beta)$ at the expected precision will have
a large impact on the accuracy of the determination of the
unitarity triangle and of the CKM parameters.
In order to show this more explicitly we take $\sin(2\beta)=0.60\pm0.06,~
\sin(2\alpha)=0.10\pm 0.10$ and the values of $\mid V_{cb}\mid,~x_d$
and $m_t$ of the vision (\ref{210}) to find \cite{BLO}
\begin{equation}\label{15}
\begin{array}{rclrcl}
\mid V_{td}\mid & = & (8.8\pm 0.4)\cdot 10^{-3}&
x_s & = & 16.3\pm 1.3 \\
Br(K^+\to \pi^+\nu\bar\nu) & = & (1.01\pm 0.11)\cdot 10^{-10} &
Br(K_L\to \pi^\circ\nu\bar\nu) & = & (2.7\pm 0.3)\cdot 10^{-11} \\
\end{array}
\end{equation}

The curve "superweak" in fig. 3 is the ambiguity curve of Winstein \cite{WIN}.
 If
$(\bar\varrho,\bar\eta)$ lies on this curve it is impossible to distinguish
 the
standard model from superweak models on the basis of CP-asymmetries in
neutral B-decays to CP-eigenstates. It is evident that in order to make this
distinction both $\sin(2\alpha)$ and $\sin(2\beta)$ have to be measured.

\vspace{10.3cm}
\centerline{Fig. 4}

\subsection{$\varepsilon_K$, $B_d^o - \bar{B}_d^o$ Mixing, $\sin(2\beta)$
and $\sin(2\alpha)$}
We next combine the analysis of sections 6.1 and 6.3.
In fig. 4 we show the allowed ranges for $\sin(2\alpha)$ and $\sin(2\beta)$
corresponding to the ranges I-III in (\ref{200})--(\ref{211})
and the range IV, defined in \cite{BLO}, together with the results of the
independent measurements of $\sin(2\beta)=0.60\pm 0.06$ and $\sin(2\alpha)$
of (\ref{220}).  The latter are represented by dark shaded rectangles.
The black rectangles illustrate the accuracy
($\Delta\sin(2\alpha) = \pm 0.04$, $\Delta\sin(2\beta) = \pm 0.02$)
to be expected from B-physics at Fermilab during
the Main Injector era \cite{FNAL}
and the first phase of LHC \cite{CA}.

The impact of the measurements of $\sin(2\beta)$ and $\sin(2\alpha)$ is
clearly visible on this plot.
In cases III and IV we have examples where the measurements of
$\sin(2\alpha)$ are incompatible with the predictions coming
from $\varepsilon_{K}$ and $B^o - \bar{B^o}$ mixing. This would be a
signal for a physics beyond the standard model. The measurement
of $\sin(2\alpha)$ is essential for this.
 Analogous comments apply to the exclusion of
superweak models.

\section{Precise Determinations of the CKM Matrix}
Let us finally concentrate on the decays of class I which being essentially
free from any hadronic uncertainties, stand out as ideally suited for the
determination of the CKM parameters.
We will use as inputs \cite{AB94}:
\begin{itemize}
 \item $\mid V_{us}\mid\equiv \lambda=0.2205\pm 0.0018$  determined
 in \cite{LR,DHK}.
Recent critical discussions of this determination and of the related element
$\mid V_{ud}\mid$ can be found in \cite{MAR}.
\item
$a\equiv \sin(2\alpha)$, $b\equiv \sin(2\beta)$ to be measured in future
B-physics experiments.
\item
$Br(\klpnn)$ to be measured hopefully one day at Fermilab (KAMI), KEK or
another laboratory.
\end{itemize}

Using (\ref{1}) and (\ref{bklpn}) one determines $\varrho$, $\eta$
and $\vcb$ as follows \cite{AB94}:
\begin{equation}\label{5}
\bar\varrho = 1-\bar\eta r_{+}(b)\quad ,\quad
\bar\eta=\frac{r_{-}(a)+r_{+}(b)}{1+r_{+}^2(b)}
\end{equation}
\begin{equation}\label{9}
\mid V_{cb}\mid=\lambda^2\left[ {{\sqrt{B_L}}\o{\eta X(x_t)}}\right]^{1/2}
\qquad
B_L={Br(\klpnn)\o 1.94\cdot 10^{-10}}
\end{equation}
$\varrho$ and $\eta$ is to be found from (\ref{3}) and (\ref{5}).
Here we have introduced
\begin{equation}\label{7}
r_{\pm}(z)=\frac{1}{z}(1\pm\sqrt{1-z^2})
\qquad
z=a,b
\end{equation}
Note that the factor in front of $\lambda^2$ in (\ref{9})
gives the parameter $A$ in
the Wolfenstein parametrization.
Using (\ref{xt}) we also find a useful formula
\begin{equation}\label{9b}
\mid V_{cb}\mid=39.1\cdot 10^{-3}\sqrt\frac{0.39}{\eta}
\left[ \frac{170 ~GeV}{m_t} \right]^{0.575}
\left[ \frac{Br(\klpnn)}{3\cdot 10^{-11}} \right]^{1/4}
\end{equation}
We note that the weak dependence of $\mid V_{cb}\mid$ on $Br(\klpnn)$
allows to achieve high accuracy for this CKM element even when $Br(\klpnn)$
is known within $5-10\%$ accuracy.
There exist  other solutions for $\varrho$ and $\eta$ coming from (\ref{1}).
As shown in \cite{AB94} they can all be eliminated on the basis of the
present knowledge of the CKM matrix.

At first sight it is probably surprising that we use a rare K-meson
decay to determine $\mid V_{cb}\mid$. The natural place to do this
are of course tree level B-decays. On the other hand using unitarity
and the Wolfenstein parametrization with $\mid V_{cb}\mid=A\lambda^2$
it is clear that $\mid V_{cb}\mid$ gives the overall scale $A$ of the
top quark couplings $V_{td}$ and $V_{ts}$ which are the only CKM
couplings in $\klpnn$. From this point of view it is very natural
to measure the parameter $A$ in a short distance process involving
the top quark and using unitarity of the CKM matrix to find the
value of $\mid V_{cb}\mid$. Moreover this strategy, in contrast to
tree-level B-decays, is free from hadronic uncertainties. On the
other hand one should keep in mind that this method contains the
uncertainty from the physics beyond the standard model which could
contribute to short distance processes like $\klpnn$. We will return
to this below.

As illustrative examples, let us consider the following three scenarios:

\smallskip
\underline{Scenario I}
\begin{equation}\label{210a}
\begin{array}{rclrcl}
\sin(2\alpha) & = &  0.40 \pm 0.08 &
\sin(2\beta)  & = & 0.70 \pm 0.06 \\
Br(\klpnn) & = & (3.0 \pm 0.3)\cdot 10^{-11} &
m_t & = & (170 \pm 5)~GeV \\
\end{array}
\end{equation}

\smallskip
\underline{Scenario II}
\begin{equation}\label{211a}
\begin{array}{rclrcl}
\sin(2\alpha) & = &  0.40 \pm 0.04 &
\sin(2\beta)  & = & 0.70 \pm 0.02 \\
Br(\klpnn) & = & (3.00 \pm 0.15)\cdot 10^{-11} &
m_t & = & (170 \pm 3)~GeV \\
\end{array}
\end{equation}

\smallskip
\underline{Scenario III}
\begin{equation}\label{212}
\begin{array}{rclrcl}
\sin(2\alpha) & = &  0.40 \pm 0.02 &
\sin(2\beta)  & = & 0.70 \pm 0.01 \\
Br(\klpnn) & = & (3.00 \pm 0.15)\cdot 10^{-11} &
m_t & = & (170 \pm 3)~GeV \\
\end{array}
\end{equation}
The accuracy in the scenario I should be achieved at B-factories
\cite{BAB}, HERA-B \cite{AL},
at KAMI \cite{AR} and at KEK \cite{ISS}.
 Scenarios II and
III correspond to B-physics at Fermilab during the Main Injector
era \cite{FNAL} and at LHC
\cite{CA} respectively.
At that time an improved measurement of $Br(\klpnn)$ should be aimed for.

The results that would be obtained in these scenarios for $\varrho$, $\eta$,
$R_t$, $\mid V_{cb}\mid$, $\mid V_{ub}/V_{cb}\mid$, $\mid V_{td}\mid$,
$\mid V_{ts}\mid$ and $\sin(2\gamma)$ are collected in table 1.
\begin{table}
\begin{center}
\begin{tabular}{|c||c||c|c|c|}\hline
& Central &$I$&$II$&$III$\\ \hline
$\varrho$ &$0.072$ &$\pm 0.040$&$\pm 0.016$ &$\pm 0.008$\\ \hline
$\eta$ &$0.389$ &$\pm 0.044$ &$\pm 0.016$&$\pm 0.008$ \\ \hline
$R_t$ &$1.004$ &$\pm 0.025$ &$\pm 0.012$&$\pm 0.006$ \\ \hline
$\mid V_{cb}\mid/10^{-3}$ &$39.2$ &$\pm 3.9$ &$\pm 1.7$&$\pm 1.3$\\ \hline
$\mid V_{ub}/V_{cb}\mid$ &$0.087$ &$\pm 0.010$ &$\pm 0.003$&$\pm 0.002$
 \\ \hline
$\mid V_{td}\mid/10^{-3}$ &$8.7$ &$\pm 0.9$ &$\pm 0.4$ &$\pm 0.3$ \\
 \hline
$\mid V_{ts}\mid/10^{-3}$ &$38.4$ &$\pm 3.8$ &$\pm 1.7$&$\pm 1.3$ \\ \hline
$\sin(2\gamma)$ &$0.35$ &$\pm 0.15$ &$\pm 0.07$&$\pm 0.04 $ \\ \hline
$B_K$ &$0.83$ &$\pm 0.17$ &$\pm 0.07$&$\pm 0.06$ \\ \hline
$F_B\sqrt{B_B}$ &$200$ &$\pm 19$ &$\pm 8$&$\pm 6$ \\ \hline
\end{tabular}
\end{center}
\centerline{}
{\bf Table 1:} Determinations of various parameters in scenarios I-III
\cite{AB94}
\end{table}

Table 1 shows very clearly the potential of CP asymmetries
in B-decays and of $\klpnn$ in the determination of CKM parameters.
It should be stressed that this high accuracy is not only achieved
because of our assumptions about future experimental errors in the
scenarios considered, but also because $\sin(2\alpha)$ is a
very sensitive function of $\varrho$ and $\eta$ \cite{BLO},
$Br(\klpnn)$
depends strongly on $\mid V_{cb}\mid$ and most importantly because
of the clean character of the quantities considered.

In table 1 we have also shown the values of the non-perturbative
parameters $B_K$ and  $F_B \sqrt{B_B}$ (in MeV) which can be
extracted from the data on $\varepsilon_K$ and
$B^{\circ}_d-\bar B^{\circ}_d$ mixing once the CKM parameters have been
determined in the scenarios considered. To this end
$x_d=0.72$ and $\tau(B)=1.5~ps$ have been assumed. The errors
on these two quantities should be negligible at the end of this
millennium. Note that the resulting central values for $B_K$ in table 1
are close to the lattice \cite{SHARP}
results.
Similar patterns of uncertainties
emerge for different central input parameters \cite{AB94}.

It is instructive to investigate whether the use of
$\kpnn$ instead of $\klpnn$ would also give interesting results for
$V_{cb}$ and $V_{td}$.
 We again consider scenarios I-III with
$Br(\kpnn)= (1.0\pm 0.1)\cdot 10^{-10}$ for the scenario I and
$Br(\kpnn)= (1.0\pm 0.05)\cdot 10^{-10}$ for scenarios II and III
in place of $Br(\klpnn)$ with all other input parameters unchanged.
An analytic formula for $\vcb$ can be found in \cite{AB94}.
The results for $\varrho$, $\eta$, $R_t$, $\mid V_{ub}/V_{cb}\mid$ and
$\sin(2\gamma)$ remain of course unchanged. In table 2 we show the
results for $\mid V_{cb} \mid$, $\mid V_{td}\mid$ and $F_B\sqrt{B_B}$.
We observe that
due to the uncertainties present in the charm contribution to
$\kpnn$, which was absent in $\klpnn$, the determinations of
 $\mid V_{cb}\mid$,
 $\mid V_{td}\mid$ and $F_B\sqrt{B_B}$ are less accurate.
 If the uncertainties due to the charm mass
and $\Lambda_{\overline{MS}}$ are removed one day, only the uncertainty
related to $\mu$ will remain in $P_0(K^+)$ giving
$\Delta P_0(K^+)=\pm 0.03$ \cite{BB3}.
 In this case the results in parentheses
in table 2 would be found.

\begin{table}
\begin{center}
\begin{tabular}{|c||c||c|c|c|}\hline
& Central &$I$&$II$&$III$\\ \hline
$\mid V_{cb}\mid/10^{-3}$ &$41.2$ &$\pm 4.3~(3.2)$ &$\pm 3.0~(1.9)$&
$\pm 2.8~(1.8)$\\ \hline
$\mid V_{td}\mid/10^{-3}$ &$9.1$ &$\pm 0.9~(0.7)$ &$\pm 0.6~(0.4)$&
$\pm 0.6~(0.4)$ \\
 \hline
$F_B\sqrt{B_B}$ &$190$ &$\pm 17~(12)$ &$\pm 12~(7)$&$\pm 12~(7)$ \\ \hline
\end{tabular}
\end{center}
\centerline{}
{\bf Table 2:} Determinations of various parameters in scenarios I-III
using $\kpnn$ instead of $\klpnn$. The values in parentheses show
the situation when the uncertainties in $m_c$ and $\Lambda_{\overline{MS}}$
are not included.
\end{table}
To summarize
we have seen that the measurements of the CP asymmetries in neutral
B-decays together with a measurement of $Br(K_L\to \pi^\circ\nu\bar\nu)$
and the known value of $\mid V_{us}\mid$ offer a precise determination
of {\it all} elements
of the Cabibbo-Kobayashi-Maskawa matrix essentially without
any hadronic uncertainties.
$\klpnn$ proceeds almost entirely through direct CP violation and is known
to be a very useful decay for the determination of $\eta$. However due
to the strong dependence on $V_{cb}$ this determination cannot fully
compete with the one which can be achieved using CP asymmetries in
B-decays. As the analysis of \cite{BB4} shows (see section 6.2) it will
be difficult
to reach $\Delta \eta=\pm 0.03$ this way if $\mid V_{cb}\mid$ is determined
in tree level B-decays. Our strategy then is to find $\eta$ from
CP asymmetries in B decays and use $\klpnn$ to determine $\mid V_{cb}\mid$.
To our knowledge no other decay can determine $\mid V_{cb}\mid$ as
cleanly as this one.

We believe that the strategy presented in \cite{AB94} is the theoretically
 cleanest
way to establish the precise values of the CKM parameters. The quantities
of class II
are also theoretically rather clean and are useful in this respect
but they cannot compete with the quantities of class I
 considered here (see our remarks in section 5.2 and in \cite{AB94}).
On the other hand once $\varrho$, $\eta$ and $\mid V_{cb}\mid$ (or A) have
been precisely determined as discussed here, it is clear that $x_d/x_s$,
$Br(\kpnn)$ and $\Delta_{LR}$ can be rather accurately predicted and
confronted with future experimental data. Such confrontations would
offer excellent tests of the standard model and could possibly give
signs of a new physics beyond it.

Of particular interest will also be the comparison of $\mid V_{cb}\mid$
determined as suggested here with the value of this CKM element extracted
from tree level semi-leptonic  B-decays \cite{STONE,BALL}.
 Since in contrast to
$\klpnn$, the tree-level decays are to an excellent approximation
insensitive to any new physics contributions from very high energy scales,
the comparison of these two determinations of $\mid V_{cb}\mid$ would
be a good test of the standard model and of a possible physics
beyond it. Also the values of $\mid V_{ub}/V_{cb}\mid$ from tree-level
B-decays, which are subject to hadronic uncertainies  larger
than in the case of $V_{cb}$, when compared with the clean determinations
suggested here could teach us about the reliability of non-perturbative
methods. The same applies to the quantities like $x_d$ and the CP violating
parameter $\varepsilon_K$ which are subject to uncertainties present in
the non-perturbative parameters $F_B \sqrt{B_B}$ and $B_K$ respectively.

It is also clear that once the accuracy for CKM parameters presented
here has been attained, also detailed tests of proposed schemes
for quark matrices \cite{HALL,ROSS} will be possible.

Precise determinations of all CKM parameters without hadronic uncertainties
 along the lines suggested
here can only be realized if the measurements of CP asymmetries in
B-decays and the measurements of $Br(\klpnn)$ and $Br(\kpnn)$
can reach the desired accuracy.
All efforts should be made to achieve this goal.
\section{Final Remarks}
In this review we have discussed the most interesting quantities which
when measured should have important impact on our understanding of the CP
violation and of the quark mixing. We have discussed both CP violating and
CP conserving loop induced decays because in the standard model CP violation
and quark mixing are closely related.

In this short review we have concentrated on the CP violation in the
standard model. The structure of CP violation in extensions of the
standard model could deviate from this picture \cite{NQ,WEIN}.
Consequently the situation in this field could turn out to be very different
from the one presented here.
 Unfortunately
in these extensions new parameters appear and a quantitative analysis of
CP violation is more difficult. The charm meson decays could turn out
to be a very good place to look for new physics effects.

Although the search for the unitarity triangle and the tests
of the Kobayashi-Maskawa picture of CP violation is
an important target of particle physics, we should not forget that what
we are really after is the true origin of CP violation observed in nature.
The strategies presented here may shed some light in which direction we
should go. However simply finding the values of $\varrho$ and $\eta$ or
demonstrating that the KM picture of CP violation is correct or false
is certainly
and fortunately not the whole story. In order to understand the true origin
of CP violation in nature we need new experiments and new theoretical
ideas.

\vskip0.5cm
I would like to thank the organizers, in particular
Prof. G. Martinelli, Prof. Diambrini-Palazzi and Prof. L. Zanello
for such a pleasant atmosphere during this symposium.
I would also like to thank all my collaborators for exciting time we
had together and Gerhard Buchalla, Robert Fleischer and Mikolaj Misiak
for a critical reading of the manuscript.

\vfill\eject

\end{document}